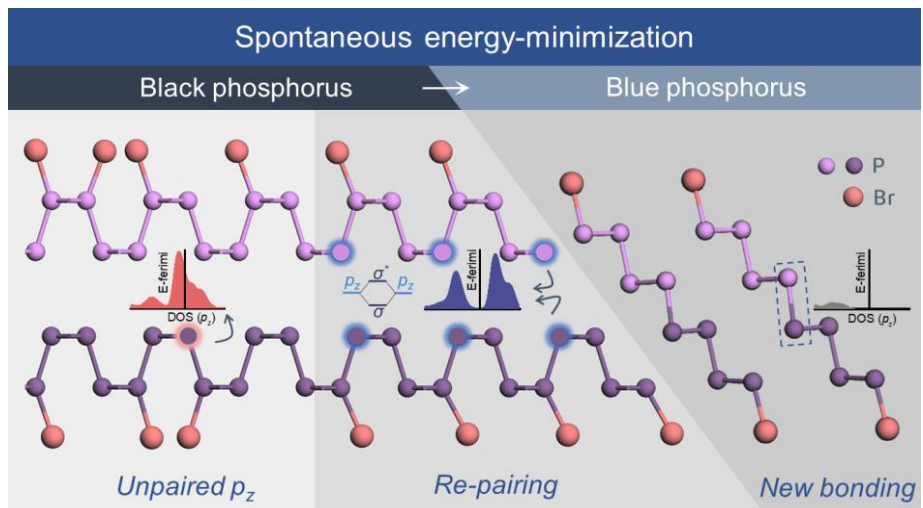

For Table of Contents Only

# Prediction of topotactic transition from black to blue phosphorus induced by surface Br adsorption

Hao Tian,[1, 2, #] Wenjun Xie,[1, #] Maohai Xie,[3] Chuanhui Zhu,[1, 2,] Hu Xu,[1, *] Shuk-Yin Tong [1, 2, 4, *]

[1]School of Science and Engineering, The Chinese University of Hong Kong, Shenzhen 518172, China

[2]Department of Physics, Southern University of Science and Technology, Shenzhen 518055, China

[3]Department of Physics, Hong Kong University, Hong Kong, China

[4]Institute of Materials Science and Devices, Suzhou University of Science and Technology, Suzhou 215009, China

*Correspondence:

zhuchuanhui@cuhk.edu.cn, xuh@sustech.edu.cn, tongsy@cuhk.edu.cn

# These authors contributed equally.

**Abstract:** Based on first-principles calculations, we propose a potential access to the yet unrealized freestanding blue phosphorus (blueP) through transformation of black phosphorus (blackP) induced by surface bromine (Br) adsorption. Formation of the Br-P bonds disrupts the original $sp^3$ configurations in blackP, generates unpaired $p_z$ electrons and induces a structural transformation that results in blueP formation by re-pairing the $p_z$ orbitals. *Ab* initio molecular dynamics simulations confirm that randomly adsorbed Br adatoms on bilayer blackP spontaneously diffuse into specific patterns to render the emergence of the blueP phase. The expected obtainment Br-passivated blueP nanoribbons exhibit tunable band gaps in a wide range and high carrier mobilities of the order of 1000 $cm^2V^{-1}s^{-1}$. This study provides an opportunity to fabricate blueP through the conversion from blackP by tuning its surface chemistry.

**Introduction:** The great interests in layered-phosphorus stem not only from their versatile properties, such as that shown in black phosphorus (blackP),[1-6] promising an array of applications, but also from their intriguing polymorphism rooted in the $sp^3$ hybridization. Unlike the $sp^2$ hybridization flattening graphite, $sp^3$ hybridization endows phosphorus atoms with the freedom of forming numerous buckled and layered allotropes[7-14] Most of them, including blue phosphorus (blueP),[7] are found to share tunable bandgaps and high carrier mobilities, which is desired for nanoelectronic devices. Tomanek *et al.*[8, 15] further put forward a wider prospect of connecting nanoribbons or patches of different phosphorus phases with distinct properties at nearly-zero energy cost so as to assemble stable in-plane heterojunctions, faceted nanotubes and faceted fullerenes. This provides exclusive opportunities to engineer low-dimensional architectures and to expand the application range of layered-phosphorus. Given the merit provided by individual phases and their combinations, the experimental realization of layered-phosphorus family is highly sought after. To our knowledge, freestanding layered-phosphorus other than blackP have however not yet been realized.

In fabricating layered-phosphorus, blueP has been under the spotlight for its attractive properties, high thermal stability,[7] and wide potentials in nanoelectronic devices,[16] photodetectors,[17] energy storage,[18] gas sensors[19] and superconductor.[20] Epitaxial growths of phosphorous were extensively studied on Au(111) with the consistent outcome of a hexapetalous-flower-like structure having the periodicity of (5×5).[21-28] Such a (5×5) structure was initially suggested to be monolayer blueP, but no atomic model was found[21, 24, 26] to satisfactorily match the experimental observations in all aspects. Recently, by combining first-principles calculations and experiments, we resolved such inconsistencies by identifying the (5×5) structure as a metallic gold-phosphorus network with strong supports lent from quantitative low-energy electron diffraction (LEED) analyses.[29] BlueP with extended dimensions was not yet realized on pristine Au(111) due to strong interactions between the substrate and phosphorus.[30] While passivated metal surfaces are recently considered to enable the fabrication of monolayer blueP,[31-32] isolating the strongly-bonded epilayer to develop application potentials remains a challenge.

In previous epitaxial growths, phosphorus source (e.g. blackP) is evaporated into single atoms which then nucleates and crystallizes on a given substrate. The phase difference between blackP and the product is delivered by a rearrangement of every single atom. Hybridized with $sp^3$, blackP and blueP in fact share a subunit consisting of a group of atoms, which could bridge the two phases with a shortcut. Based on first-principles calculations (see the calculation details in the supporting information), here we propose a transformation from blackP to blueP through recombining the shared subunits. The required assistance to this transformation is surface Br adsorption on blackP, and the avoidance of using substrate enables the freestanding blueP. Taking the bilayer blackP as the prototype, we begin by elaborating the geometry transformation trajectory from blackP to blueP with the shared subunit. The thermodynamic origin for surface Br adatoms to drive the transformation is then revealed. The mechanism of geometry transformation is confirmed by using *ab initio* molecular dynamics (AIMD) simulations. The expected products are freestanding Br-passivated zigzag blueP nanoribbons (BrPNRs), which inherits high carrier mobility and tunable bandgaps of blackP, highlighting the potential in electronic applications. The result of this study thus not only provides an opportunity to fabricate blueP through the conversion from blackP by tuning its surface chemistry, but also deepens an understanding of adsorbate-substrate interaction where adsorbates "catalyze" the behavior of substrate.

**Results and Discussion:** The $sp^3$ hybridization in layered-phosphorus makes blackP and blueP to have shared subunits and comparable bulk energies. The first prediction of blueP structure was inspired from dislocation of blackP's subunit.[7] In principle, these features would favor transformations between blackP and blueP as well as between other allotropes (see **Fig. S1 and S2** in the supporting information). Geometrically, the blueprint for accessing blueP through blackP has two steps, a layer shift (**Fig. 1a**) and a shear deformation (**Fig. 1b**). In **Fig. 1a**, we define the AD stacking of blackP by antiparallelly shifting two adjacent layers stacked in the ground state (i.e., the AB stacking) for a relative displacement of a/4 along [100] and b/2 along [010] direction. Such an AD stacking is readily achievable, as it is one of the two locally stable configurations in the energy landscape (see **Fig. 1c**) with only 3.8 meV/atom energy increment over the AB staking.

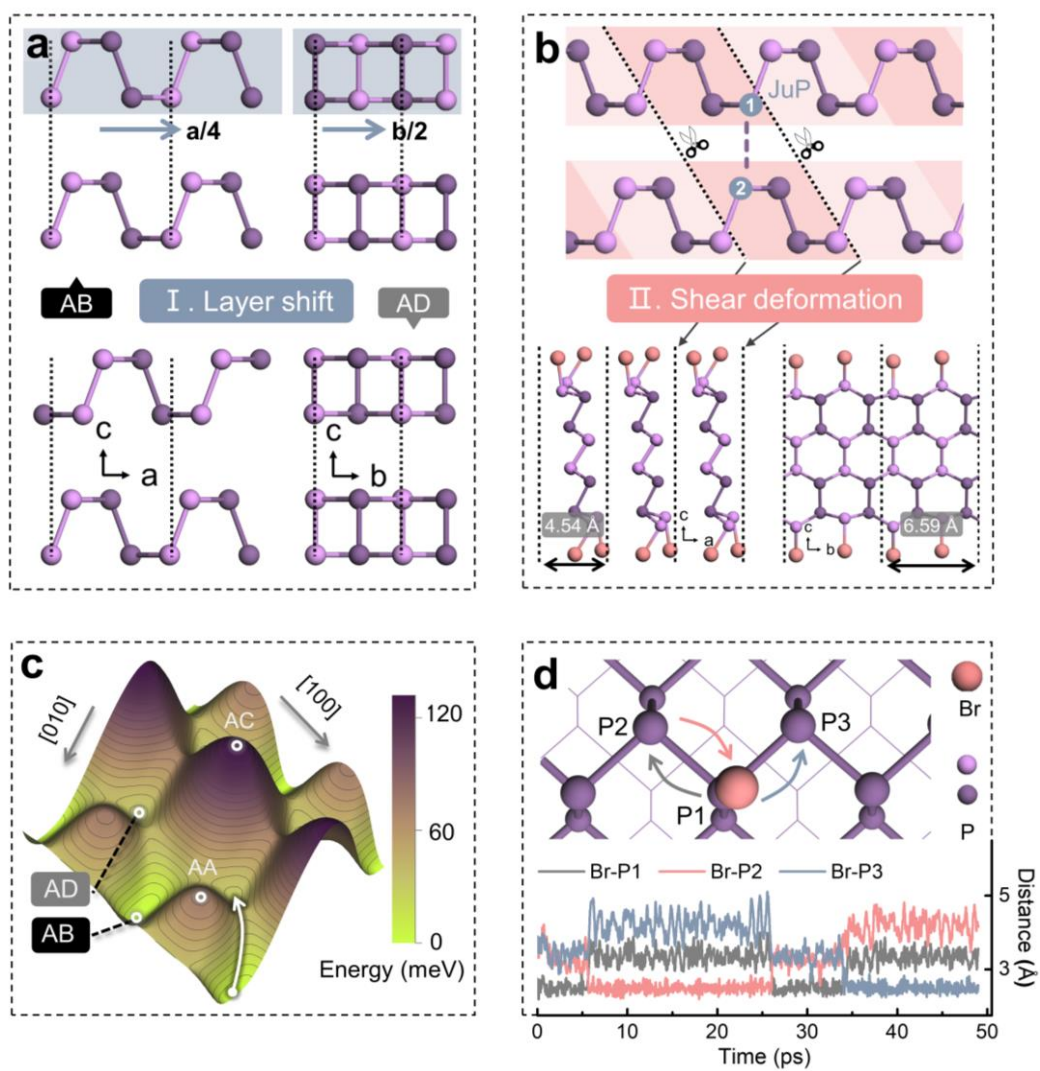

**Figure 1. Geometry transformation from blackP to blueP.** (a) Antiparallel shuffling of two adjacent layers in the AB-stacked blackP for a relative displacement of a/4 along [100] and b/2 along [010] to achieve the AD stacking. (b) Scission of chemical bonds between zigzag chains. The rhomboid patches mark the phosphorus subunits. Both dark and light purple spheres are P. The pink ones represent Br. (c) Energy landscape (per unit cell with 8 P atoms) of bilayer blackP with different stacking orders. The top layer is moved with a step size of ~0.1 Å along the [010] and the [100] directions to map the stacking energy. The white arrow indicates a pathway

of transition from the ground AB stacking to the metastable AD configuration. The energy of the AB stacking is set to zero. (d) The distance between an adsorbed Br on blackP and the specific P atoms (depicted in top panel) as a function of AIMD simulation time. The typical length of P-Br bond is ~2.4 Å. Based on the Br-P distance variation shown in the bottom panel, we deduce a diffusion trajectory of P1 → P2 → P1 → P3 for Br in 50 ps at 400 K.

In the AD stacked blackP configuration, there are pairs of juxtaposed phosphorus atoms (named JuP atoms, e.g. those labelled by "1" and "2" in **Fig. 1b**) facing each other across the interlayer space. In the second step (**Fig. 1b**), the shared subunit (highlighted with rhomboid patches in the top panel) is isolated from the AD stacked blackP and joint into blueP by re-bonding the JuP atoms. Specifically, phosphorus subunits isolated by scissoring the existing bonds of the JuP atoms not being in parallel with the corrugated layer are rejointed by connecting JuP atoms 1 and 2 with new interlayer bonds (see the schematic illustration in the top panel of **Fig. 1b**). With this shear deformation, one finds newborn blueP ribbons with dangling bonds on their edges. In our scheme, the dangling bonds happen to be statured by the 'transformation trigger' (Br adatoms), preventing the reverse process and giving rise to stable BrPNRs structure as shown in the bottom panel of **Fig. 1b**. A crucial reason we choose Br atoms as the transformation trigger is because they may diffuse on blackP surfaces to find the desired 'trigger sites' rather than stick on the surface randomly. AIMD simulations indicate that an adsorbed Br atom on blackP changes its anchor site three times in 50 ps at 400 K (**Fig. 1d**), as benefited from the small diffusion barrier (< 0.2 eV) along the zigzag chain (see **Fig. S3**). The interesting finding of this work is that Br adatoms on blackP would diffuse into a specific pattern to induce the mentioned two-step blackP-to-blueP transformation.

To illustrate the thermodynamic drive for the Br-induced blackP-to-blueP transformation, we start with the bonding in blackP. The $sp^3$ hybridization governing both blackP and blueP structures can be visualized by the electron localization function (ELF) isosurface, and **Fig. 2a** presents one for an isovalue of 0.83. Three electrons form three covalent σ-bonds with neighboring atoms, while the remaining two occupy a lone pair orbital as expected from the valence shell electron pair repulsion model.[33] In other panels of **Fig. 2**, we show results for a slightly higher isovalue of 0.91 for interest of clarity, where the lone pairs (yellow coloured) become more prominent. As shown in **Fig. 2b**, the essential ingredient of blackP-to-blueP transformation is the $sp^3$ reorientation (or lone pair reorientation) of JuP atoms located at the boundary of subunits. As illustrated, the original lone pair gives way to a new interlayer bond while a new lone pair is created at the otherwise broken bond location, namely a reconstructing reoriented $sp^3$. The dashed lines in Fig. 2b depict the correspondence between JuP atoms involved before and after going through such a $sp^3$ reorientation.

To render this $sp^3$ reorientation, an external force is invited to break one existing bond so as to adapt the original $sp^3$. Br adsorption could serve this purpose. The lone pair on each P of blackP surface attracts electron acceptors such as halogen (see panel i of **Fig. 2c**). Br adatoms, for example, can be generated from the dissociation of $Br_2$ molecules (see **Figs. S4-6 and the following discussion**). The adsorption of a Br atom takes ~0.30 e from the surface according to the Bader charge calculation and establishes a Br-P bond, which immediately repel the original $sp^3$ configuration by impairing the lone pair (see panel ii of **Fig. 2c**). The disturbance to $sp^3$ also stretches the P-P bond from 2.257 Å to 2.428 Å between the Br-anchored site and the distal site facing the interlayer space (see panel i and ii of **Fig. 2c**). This consequently unpairs part of the $p_z$ electrons of the distal phosphorus atom (corresponding to JuP atom afte in AD stacking), as evidenced by the gap state born in the projected density of states (DOS) (panel ii of **Fig. 2d**). The yet unpaired $p_z$ electrons can then pair up with another unpaired $p_z$ electrons nearby and generated from the opposite layer by the same Br adsorption (panel iii of **Fig. 2c**). These $p_z$ orbitals tend to split into a bonding orbital σ and an antibonding orbital σ* with a reduced energy as indicated by the projected DOS shown in panel iii of **Fig. 2d**. The ELF map shown in **Fig. 2e** also captures this bonding feature, where localized electrons emerge at the interlayer space and connect the two facing JuP atoms after a dual-surface-adsorption of Br. Note that, in such a AB stacking (panel iii of **Fig. 2c**), interlayer chemical bonding has not accomplished and is hindered mainly by the large inter-atom distance of ~3.6 Å compared with a typical P-P bond of lengths of 2.2~2.3 Å. Nevertheless, simply shifting AB-stacked blackP into AD stacking could achieve a distance reduction of 0.45 Å and, remarkably, also a favourable geometry for an interlayer P-P σ bond. In the AD-stacked blackP with aforementioned Br dual-surface-adsorption, interlayer bond between JuP atoms can be established with reoriented lone pairs appearing in again well-defined $sp^3$ configuration. This is illustrated in panel iv of **Fig. 2c**. Upon pairing the $p_z$ electrons, gap states become quenched as shown in panel iv of **Fig. 2d**.

The geometry favourability for interlayer bonding in the AD stacked blackP, as shown in **Fig. 1b**, is featured by its lattice resemblance with blueP, and in turn, this resemblance gives birth to blueP nanoribbons after the completion of the transformation depicted in panel iv of **Fig. 2c**. The key to driving the transformation is the energy reduction originated from the stabilization of the unpaired $p_z$ electrons aroused by Br adsorption. In blackP thicker than two layers, identical driving factor could work as well on surfaces. For the interior layers, the adjacent close-to-surface layer, once transformed by yet more close-to-surface layers, could play a similar triggering role in promoting interlayer bonding. This is demonstrated for AB-stacked trilayer blackP, showing similar interlayer bonding in the ELF map (see **Fig. S7**). With the formation of such interlayer bonds and the scission of other bonds involved in a similar way, wider zigzag nanoribbons could emerge. On the other hand, with increasing thickness, resistance arises from the energy cost for blackP-to-blueP transformation, as blackP has a slightly lower bulk energy than blueP. Considering the balance between the surface energy reduction and the thickness-dependent energy cost, we build a linear model and estimate that the transition is allowable up to 5 layers (~20 Å) of blackP (see **Fig. S8** and **Table S1**).

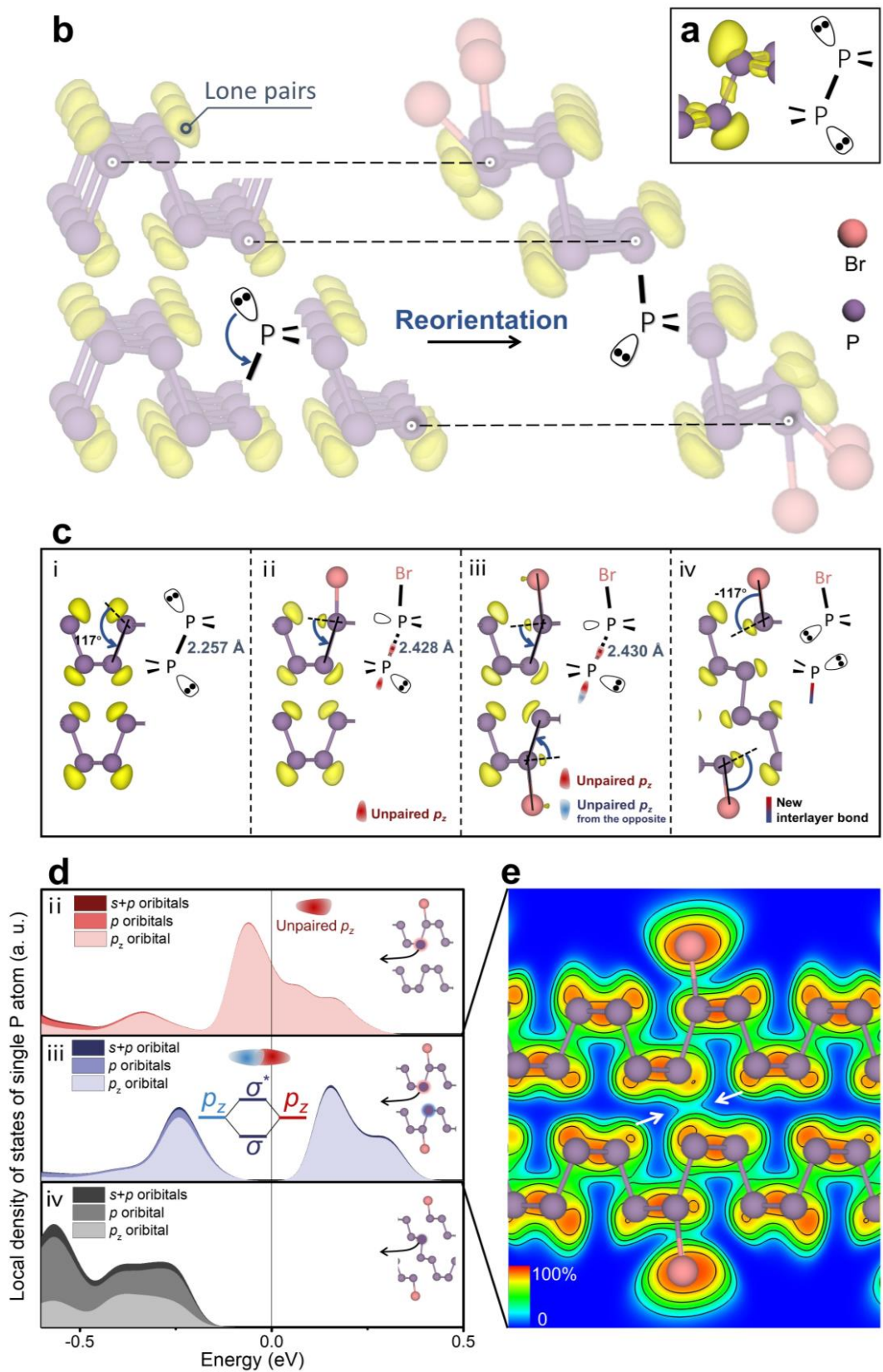

**Figure 2. Thermodynamic driving force for blackP-to-blueP transformation.** (a) ELF isosurface (at isovalue of 0.83) and corresponding schematic depiction of $sp^3$ hybridization in layered-phosphorus. (b) Atomic structures of AB-stacked blackP and Br-passivated blueP zigzag nanoribbons with ELF isosurface (at isovalue of 0.91),

showing $sp^3$ reorientations. (c) Lone pairs in (i) pristine AB-stacked blackP, (ii) AB-stacked blackP with a Br adatom, (iii) that with dual-surface Br adsorption, and (iv) AD-stacked blackP with dual-surface Br adsorption. (d) Local density of states projected on different orbitals of the JuP facing the interlayer space after Br adsorption in circumstances (ii)-(iv), corresponding respectively to that described in (c). For (ii) and (iii), the gap states are mainly contributed from the $p_z$ orbital. (e) ELF map of bilayer blackP with dual-surface Br adsorption. The scale bar for ELF value is given in the bottom-left corner.

The dual-surface-adsorption depicted in panel iii of **Fig. 2c** promoting interlayer bonding requires Br adatoms to anchor on specific sites. This can be seen more clearly in **Fig. 2e** where one adsorbed Br atom should match the other on the opposite surface such that the resulted lone $p_z$ electrons could pair with each other across the interlayer space. Using AIMD simulations, the surface dynamics of randomly distributed Br adatoms is investigated at 400 K. We find that those favourable sites are indeed what Br adatoms seek by diffusion and driven by energy minimization.

Instead of a fully random distribution, we start the simulation with 1/4 ML Br adatoms arranged in the "bridge mode" on both surfaces of a (3×4) AB-stacked blackP slab (named the random bridge 1 (RB1) and depicted in the top panel of **Fig. 3a**), as "bridge mode" represents the most stable adsorption mode on the AB-stacked blackP, benefitted from the π-like bonding effect [34] (see the detailed discussion following **Fig. S4-6**). In the first 10 ps of simulation, Br adatoms are seen to vibrate around their assigned anchor sites as marked by purple and white in **Fig. 3b,** which shows statistical distributions of the adatom locations projected on the (001) plane. Facilitated by a mild temperature of 400 K, they diffuse to form a pattern in which only the white sites become populated (see **Fig. 3c** for the last 10 ps (140-150 ps) of the simulation). The snapshot for the RB1 at 150 ps is given in the bottom panel of **Fig. 3a**, where the side view of the patterned Br adsorption is seen to accord with that shown in **Fig. 2e**, promoting interlayer bonds. Given the time-dependent Br distribution shown in **Figs. 3b, 3c, 3e and 3f**, we conclude that the diffusion kinetics for the evolvement of the specific pattern is along the zigzag chains, and **Fig. 3d** shows Br position distribution on the top of blackP surface over the time span of 0-150 ps with marked diffusion pathway of adatoms.

With Br adatoms diffusing into the patterned configuration, the phosphorus structural transformation commences via layer shifts and accompanied by the interlayer space shrinkage, necessitating interlayer bonding. As seen from **Fig. 3g**, the phosphorus structural transformation for the RB1 configuration occurs at ~110 ps, as is signified by a sharp interlayer distance reduction of ~0.25 Å. At the same time, layer shifts are triggered with displacements of around 1/4 units of the basis vector along [100] of blackP and 1/2 units along [010] (see **Fig. 3h** for the time-dependent layer displacements projected on [100] and [010] directions). The successive AD stacking is reached with an energy reduction of ~12 meV per P atom, allowing the interlayer bonding. It is noted that the system energy decreases further by ~7 meV per P atom at 143 ps when the AD stacking has emerged for ~33 ps (see **Fig. 3h**). From Figs. 3e and 3f, one sees that Br adatoms have not completely achieved the desired pattern on both surfaces at 120 ps. This energy reduction is thus the result of a process where some Br adatoms continue to anchor and diffuse to the desired pattern after the layer shift. It suggests that Br adatoms of less than 1/4 ML would induce the layer shift (also see **Fig. S9** and the following discussion) and adding adatoms on AD stacked blackP would diffuse into the pattern in short time.

To test the sensitivity of the structural transformation on the initial Br adsorption pattern, we construct a different configuration called RB2 as shown in **Fig. S10**. The critical transformation time for the RB2 is slightly shortened to ~55 ps (c.f. **Fig. 3g and Fig. 3i**). Dictated by symmetry, here the AD stacking is reached via an inverted layer displacement in [010], in contrast to the case of RB1. As is clearly indicated by the results of **Fig. 3i** between 55-120 ps, newly formed interlayer bonds should lock the AD stacking for more Br adatoms getting involved. Adding an extra 1/4 ML Br, increasing the total coverage to 1/2 ML, will completely transform the bilayer blackP into Br-passivated blueP nanoribbons (c.f., **Fig. 1b)**. It can be deduced that the transformation dynamic for thicker blackP would become slower due to the increased energy barrier for layer shift and the diminished thermodynamic driving force caused by the rising energy cost for lattice conversion. For AD-stacked trilayer blackP with randomly distributed 3/8 ML Br adatoms on surfaces, it takes ~9 ps to complete the transformation, while in 50 ps, AB-stacked trilayer blackP with pre-patterned Br at the same coverage has not transformed. The layer shift is suggested to be the rate-determining step in the transformation.

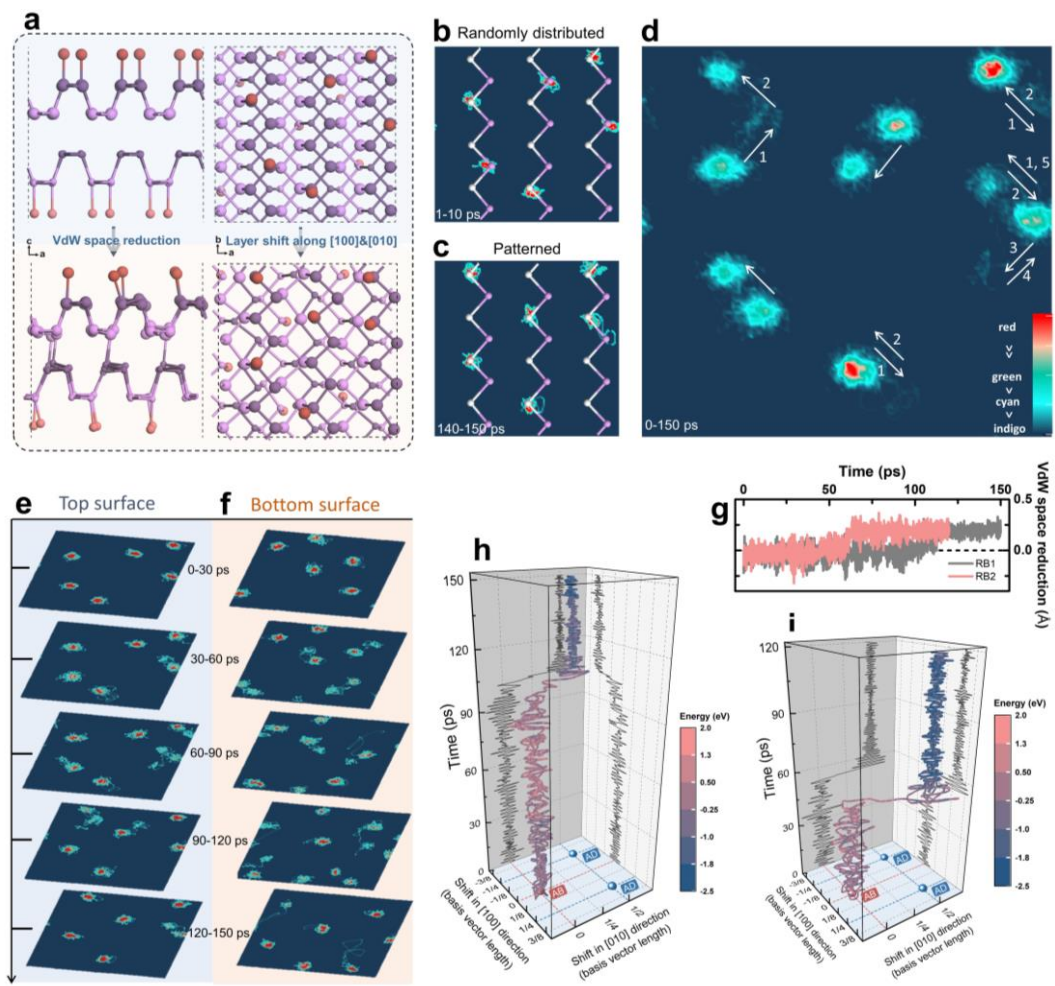

**Figure 3. Kinetics of blackP-to-blueP transformation.** (a) Atomic model of the RB1 configuration at the start (0 ps) and the end (150 ps) of the AIMD simulation. (b)-(f) Probability distributions of Br adatoms' locations projected on the (001) plane with different simulation time intervals as indicated. In the simulation, blackP layer has shifted, dragging Br adatoms on surface to follow. The real-time layer displacement has been subtracted from the Br adatoms' coordinates in the statistical analysis so as to show the relative diffusion of Br on blackP surfaces. (b) Initial 10 ps of simulation showing initial random distribution of Br adatoms. (c) The last 10 ps (i.e., from 140 to 150 ps) simulation revealing patterned Br distribution on the same surface. (d) Br distribution accumulated over the full time span of 0-150 ps. (e, f) Br

distribution on the top (e) and bottom (f) surfaces for every 30 ps of the simulation. (g) Interlayer distance reduction as a function of the simulation time. (h, i) Relative shift between adjacent layers and system energy variation for the system consisting of 96 P and 12 Br atoms as a function of the simulation time for the RB1(h) and RB2 (i) configurations, respectively. Layer displacements projected on (100) and (010) directions are also shown.

Adsorption of Br on blackP not only transforms it into a new phase, blueP, but also reduces the dimension of the structure. The transformation product is freestanding zigzag blue phosphorus nanoribbons with Br terminals on their edges (BrPNRs). BrPNRs has a rectangular cell with lattice vectors $\vec{a}$ = 4.54 Å between the interlayer and $\vec{b}$ = 6.59 Å along the zigzag direction (see the bottom panel of **Fig. 1b**). The two adjacent Br atoms experience Coulomb repulsions as shown, and the Wigner-Seitz cell contains 16 P atoms and 4 Br atoms. Calculations of phonon spectra and AIMD simulations suggest that such BrPNRs are both dynamically and thermodynamically stable at room-temperature (**see Fig. S11 and the following discussion**). Br-terminals probably play a role of self-passivation to protect stacked BrPNRs from degradation, as suggested by our calculations (**see Fig. S12 and the following discussion**). A moderate exfoliation energy of ~80 meV/atom suggests the feasibility of exfoliating layered BrPNRs.[35] Tailored by the thickness, the band gap of BrPNRs spans a wide range, covering and bridging that of blackP and blueP (**Fig. 4a**). Electronic structure calculations (shown in **Fig. 4b**) reveal that the bulk of layered BrPNRs is a semiconductor with an indirect band gap of ~0.29 eV, while it is widened to ~1.84 eV for a monolayer. In addition, the bulk of BrPNRs shows a relatively high hole mobility of 1101.36 $cm^2V^{-1}s^{-1}$ along the [100] direction (Table 1). Carrier mobility of isolated 1~3 layers of BrPNRs are given in **Table S2**. These promising properties highlight the potentials of BrPNRs in future electronic applications. The BrPNRs could also be used as initial materials for bottom-up synthesis of extended blueP monolayer through radical-coupling reactions as demonstrated in carbon chemistry.[36-37]

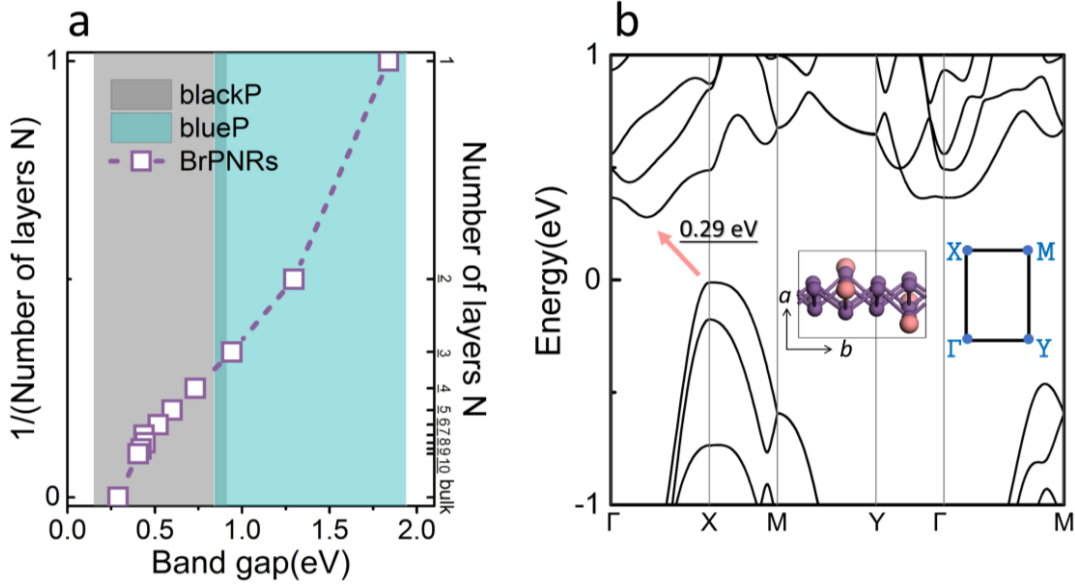

**Figure 4. Electronic properties of BrPNRs.** (a) Band structure of bulk 2D BrPNRs, together with top view of its atomic structure and the schematics of the reciprocal lattice. (b) Band gap tunability of BrPNRs in comparison with blackP and blueP. Band gaps of blackP and blueP are based on PBE level results of Qiao et al.[38] and Pontes et al.[39], respectively.

**Table 1. Predicted carrier mobility ($\mu$) of single layer BrPNR at 300K.**

| Carrier type | effective masses | | deformation potential (eV) | | 2D elastic modulus (Jm$^{-2}$) | | mobilities (cm$^2$V$^{-1}$s$^{-1}$) | |
|---|---|---|---|---|---|---|---|---|
| | $m_x^*/m_0$ | $m_y^*/m_0$ | $E_{1x}$ | $E_{1y}$ | $C_{x\_2D}$ | $C_{y\_2D}$ | $\mu_{x\_2D}$ | $\mu_{y\_2D}$ |
| e | 24.61 | 0.27 | 1.59 | 10.59 | 186.29 | 13.48 | 24.72 | 3.67 |
| h | 0.17 | 1.07 | 7.02 | 0.81 | 186.29 | 13.48 | 1101.36 | 951.04 |

**Conclusion:** In summary, we propose that surface Br adsorption on blackP surfaces can transform it into freestanding blueP nanoribbons. Formation of the Br-P bonds disrupts the original $sp^3$ configurations dominated in blackP, and generates unpaired $p_z$ electrons, inducing a structural transformation that results in blueP formation by re-pairing. AIMD simulations suggest that randomly adsorbed Br adatoms on bilayer blackP spontaneously diffuse into specific patterns to render the emergence of the blueP phase. The transformation is allowed up to five layers of blackP, and the rate-determining step is associated to the layer shift step. This work reports the opportunity to fabricate freestanding blueP nanoribbons with application potentials using massively producible blackP and suggests an unconventional surface chemistry concept to drive bulk reconstruction.

**Acknowledgements**

This work is financially supported by the Guangdong Natural Science Funds for Distinguished Young Scholars (No. 2017B030306008), the National Natural Science Foundation of China (Grants Nos. 11974160, 11674148, 11334003, 11404159 and 11704177), and the Center for Computational Science and Engineering of Southern University of Science and Technology. MHX acknowledge the support of a GRF grant (No. 17304318) from the Research Grant Council of Hong Kong Special Administrative Region, China.

**Supplementary Information**
Inducing black-to-blue phosphorus transformation by Br surface adsorption

**Calculation details**

Our results are based on first-principles calculations using Vienna ab initio simulation package (VASP)[1]. The electronic exchange correlation is described using the Perdew-Burke-Ernzerhof (PBE) form[2] of the generalized gradient approximation (GGA). The projector augmented wave method[3,4] is used to treat electron–ion interactions with a plane-wave-basis cutoff of 500 eV. The vacuum region is set to ~ 12 Å to avoid interlayer interaction, and all structures are optimized by the conjugate gradient method until the Hellmann−Feynman force is less than 0.02 eV/Å. The climbing image nudged elastic band (CI-NEB) method is used to investigate the minimum reaction pathways[5]. The AIMD calculations are performed using a (3×4) blackP slab, and a corresponding 2×2×1 Gamma centered k-points mesh is set. A temperature of 400K is simulated with a time-step of 1 fs by rescaling the velocities in each step. We employ the DFT-D3[7] method of Grimme to evaluate the van der Waals (vdW) effect in all calculations.

The carrier mobility for 2D Br-PNRs is estimated by the expression[8,9]

$$\mu_{2D} = \frac{e\hbar^3 C_{2D}}{k_B T m^* m_d (E_l^i)^2} \quad (1)$$

where $e$, $\hbar$ and $k_B$ denote respectively the charge of electron, the reduced Plank's constant and Boltzmann's constant. $T$ is temperature which is set at 300 K. $m^*$ (i.e., $m_x^*$ or $m_y^*$) represents the effective mass along the transport direction and $m_d = \sqrt{m_x^* m_y^*}$ gives the average effective mass $m_d$. The term $E_l$ denotes the deformation potential constant of the valence-band minimum for hole or conduction-band maximum for electron along the transport direction and is determined by $E_l^i = \Delta V_i / (\Delta l / l_0)$. Here $\Delta V_i$ is the energy change for the $i^{th}$ band after -1% ~ +1% cell compression and dilatation, $l_0$ is the lattice constant in the transport direction and $\Delta l$ is the deformation of $l_0$. The two-dimensional (2D) elastic modulus $C_{2D}$ is given by $(E - E_0)/S_0 = C_{2D}(\Delta l / l_0)^2 / 2$, where $E$ represents the total energy for the 2D Br-PNRs and $S_0$ is the equilibrium lattice volume. The mobility for isolated 1D Br-PNRs is [8,10]

$$\mu_{1D} = \sqrt{\frac{2}{\pi}} \frac{e\hbar^2 C_{1D}}{(k_B T)^{1/2} (m^*)^{3/2} (E_l^i)^2} \quad (2)$$

## Structural transformation of pnictogen layered materials

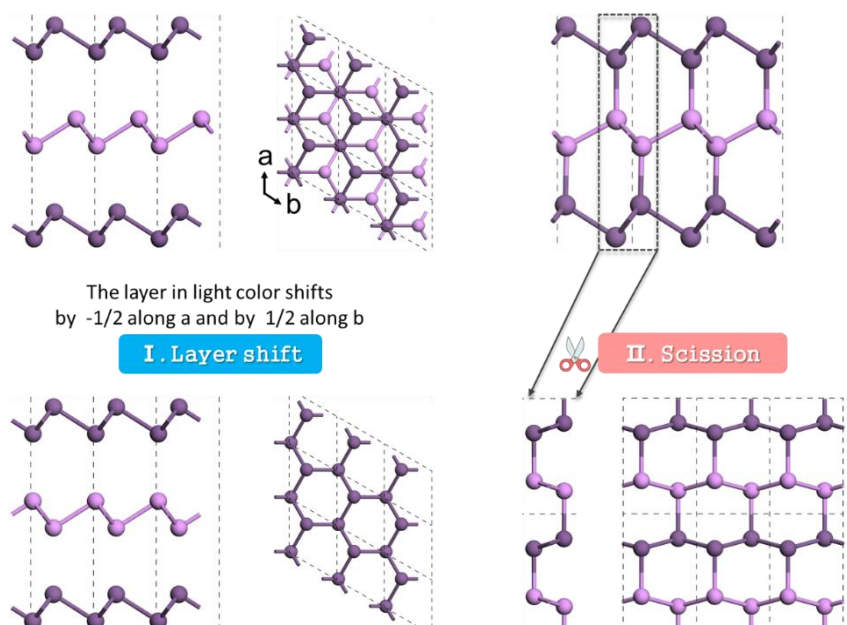

Fig. S1. Schematic illustration of transition from β-P (blue) to γ-P phosphorus by two steps.

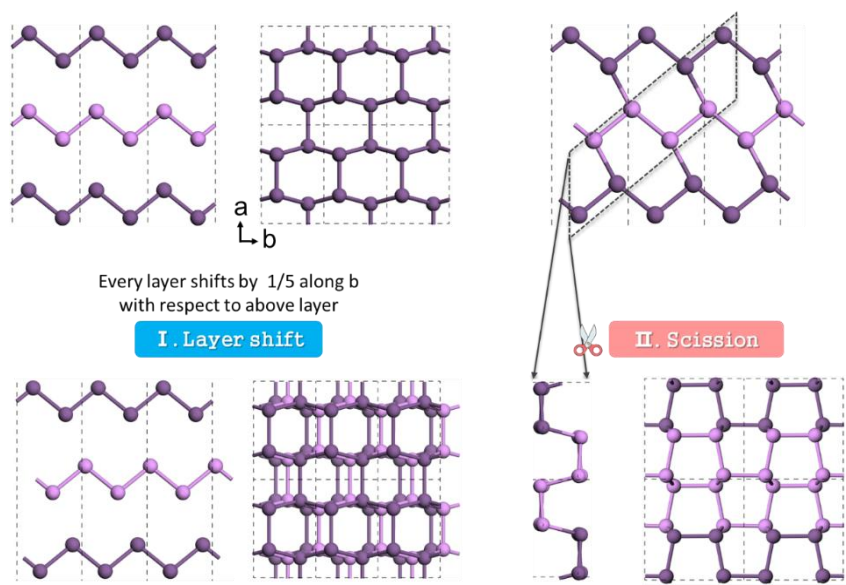

Fig. S2. Schematic illustration of transition from γ-P to δ-P phosphorus by two steps.

In addition to transition between $\alpha$ and $\beta$ phases, similar geometric transformation could also be constructed for other layered phosphorus allotropes such as the well-known γ-P and δ-P[11,12] because of the shared $sp^3$ hybridization. δ-P can be obtained from γ-P, and γ-P can be obtained from blueP (β-P) by transformations.

**Diffusion of Br adatom on black-P surface**

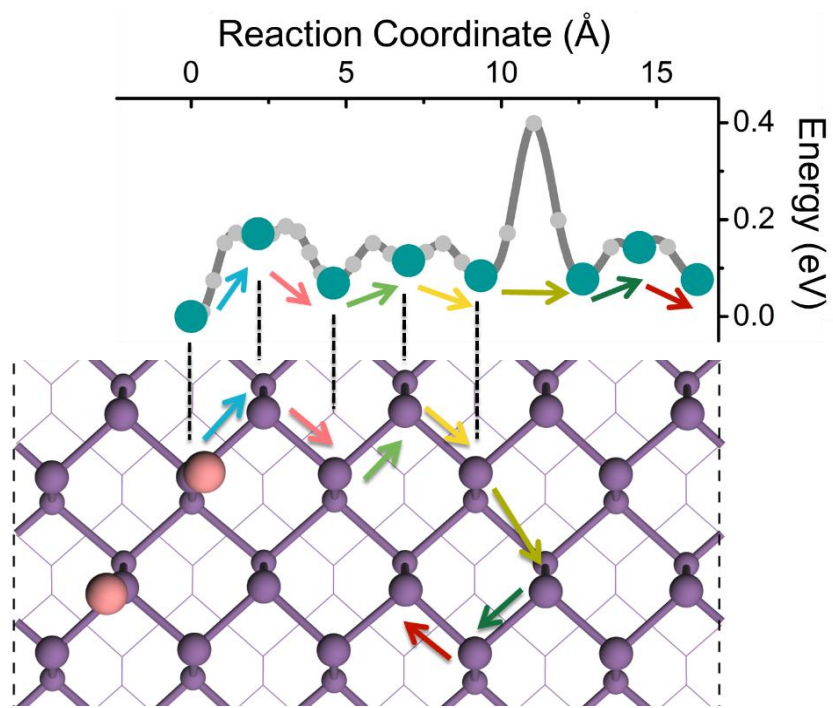

Fig. S3. Potential energy profile for Br adatom diffusion along and across the zigzag chain on blackP. For single Br adatom after $Br_2$ dissociation, the diffusion barrier along the zigzag chain is smaller than 0.2 eV.

## Dissociation of Br₂ on black-P

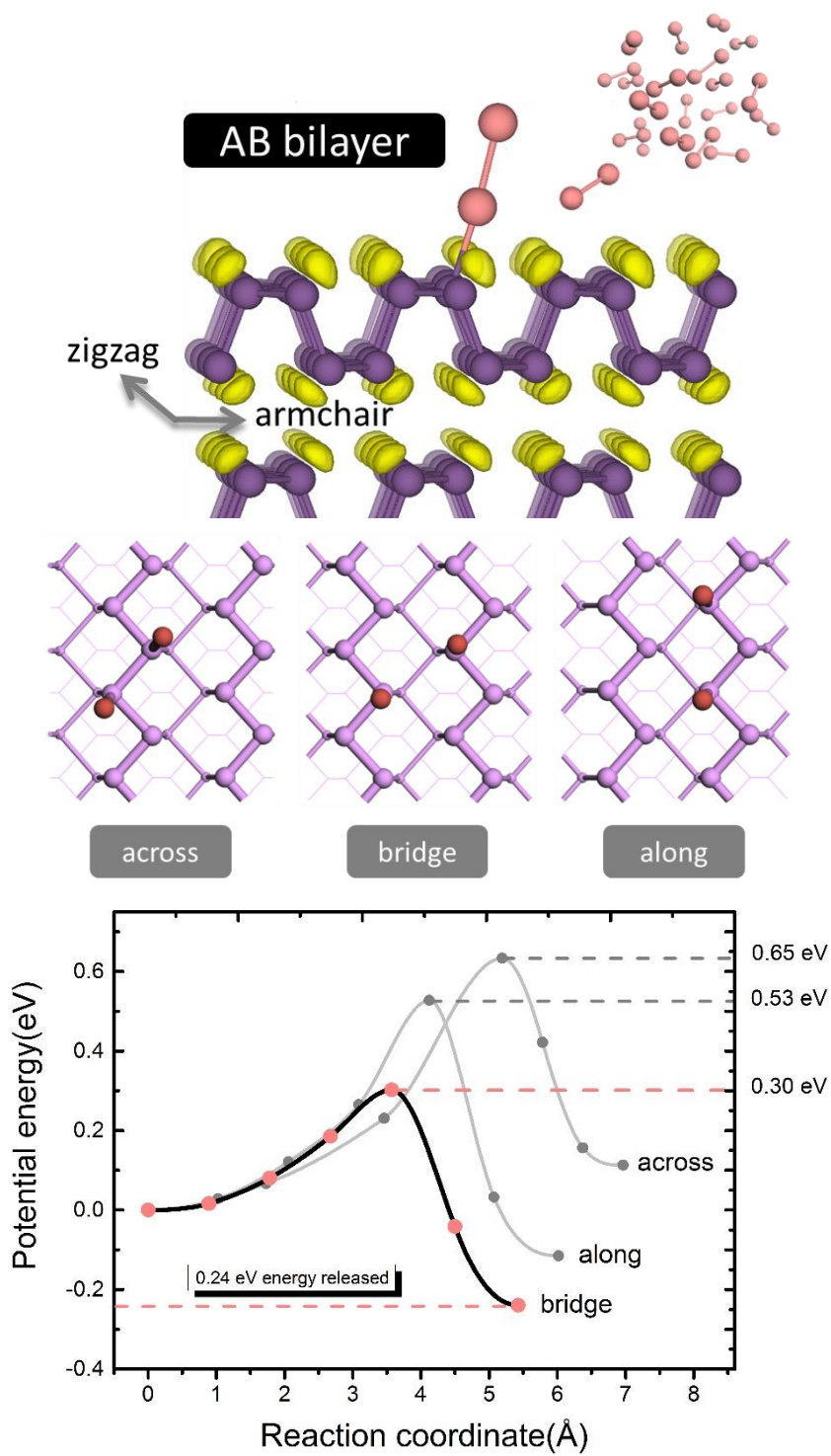

Fig. S4. Energy profiles (bottom) for conventional Br₂ dissociation in the 'across', 'along' and 'bridge' modes and their corresponding atomic structures (middle).

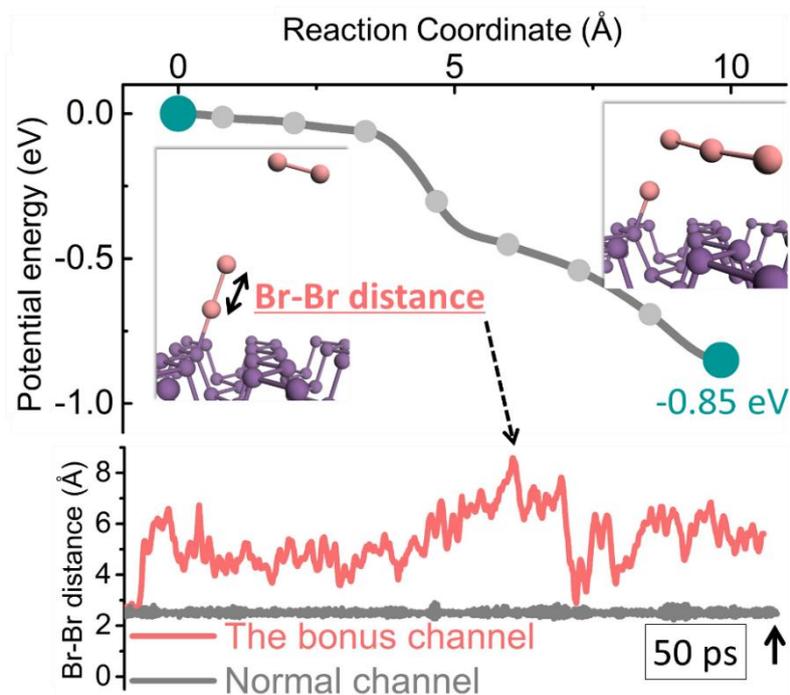

Fig. S5. Potential energy profile (above) and intramolecular Br-Br distance as a function simulation time (below) for the new dissociation mode.

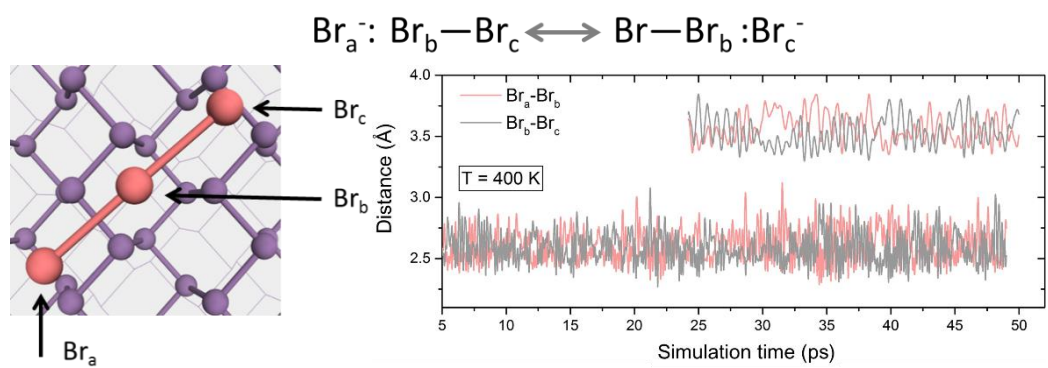

Fig. S6. Atomic structure of $Br_3^-$ on bilayer black-P and the Br-Br distance as a function of simulation time. Tribromide anion ($Br_3^-$) is a resonance hybrid exhibiting 3-center, 4-electron bonding. There are two resonance structures $Br_a^-$ : $Br_b$—$Br_c$ ↔ $Br_a$ — $Br_b$ : $Br_c^-$, where "—" denotes a single bond and ":" denotes a weak interaction. The resonance character can be identified by the variation of Br-Br distance in the MD simulation.

A Br$_2$ molecule tends to occupy the lone pair with an energy gain of 0.71 eV. There are three common dissociation modes (Fig. S4) for the adsorbed Br$_2$, in analogy to the F$_2$ dissociation on black-P.[13] The most favorable one is the 'bridge' mode (refer to dissociation configurations in middle of Fig. S4), where an energy of 0.24 eV is released after overcoming a 0.3 eV barrier. The preference of the bridge-mode dissociation is benefited from the exclusive formation of the π-like bond.[13] Bridge adsorption of Br results in $sp^3$ to $sp^2$ transition with the appearance of two adjacent $p_z$-orbitals, forming an in-plane π-like bond. This effect also explains the variations of the system energy when Br diffuses on the zigzag chain as presented in Fig. S3.

Besides the conventional dissociation modes, we also emphasize a more preferred channel due to the presence of the $d$ electrons in Br comparing to F. As shown in Fig. S5, an approaching Br$_2$ readily extract one Br atom from pre-adsorbed Br$_2$ to generate Br$_3^-$ without energy barrier, leaving behind one Br adatom on surface. The produced Br$_3^-$ is a resonance hybrid tribromide anion whose resonance characteristics can be captured in molecular dynamics (MD) simulation (Fig. S6). MD simulation shown in Fig. S5 suggests such reaction occurs in less than 1 picosecond at 400 K, whereas single Br$_2$ molecular dissociation has not been observed in 50 picoseconds. The varying Br-Br distance describes free diffusion of Br$_3^-$ on blackP surface.

Halogen was introduced to functionalize 2D materials in the way of intercalation or halogenation.[14,15] Two side fluorination has been experimentally demonstrated on graphene.[14] Experiments also show that fluorinated blackP remains stable rather than burned down to PF$_5$.[16] The procedure of exposing both surfaces of blackP sheets to elemental bromine thus seems technically feasible. Other alternative bromine sources besides bromine molecules could be XeBr$_2$ or CBr$_4$.[14,17,18]

**Interlayer bonding in trilayer blackP**

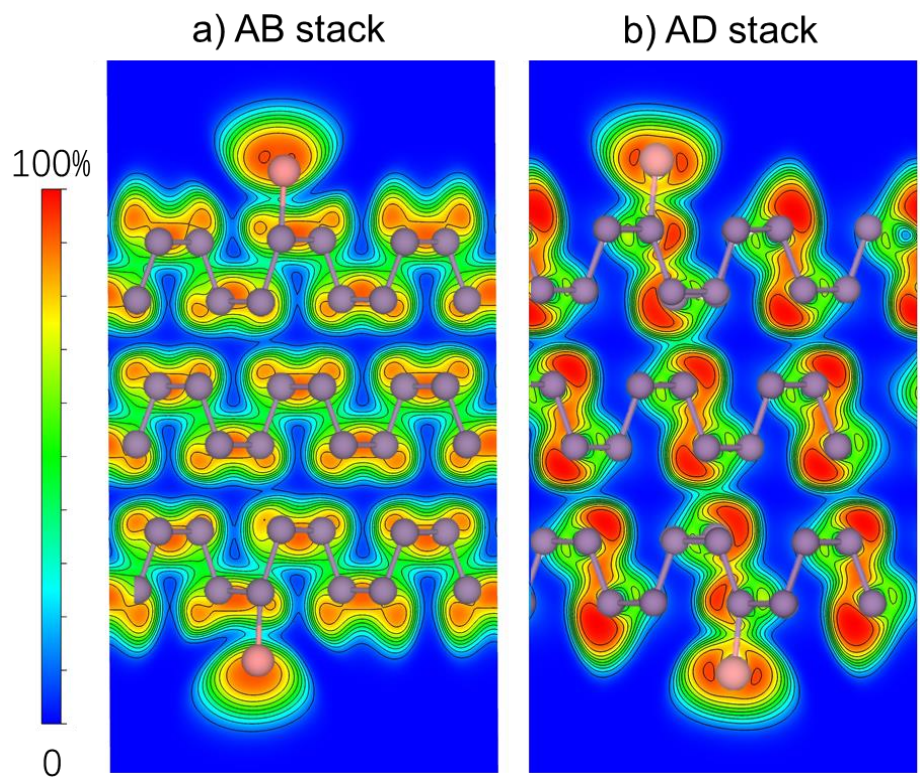

Fig. S7. Electronic localization function (ELF) maps plotted with 100% of the peak amplitude value for (a) dual-surface-Br-adsorption on AB stacked trilayer blackP and (b) with AD stack. Patterns shown in (a) and (b) are distinct, because different slices of {010} group are chosen for AB and AD stacking configurations in order to reveal the interlayer bonding feature in each clearly.

**Linear model for blackP-to-blueP transformation**

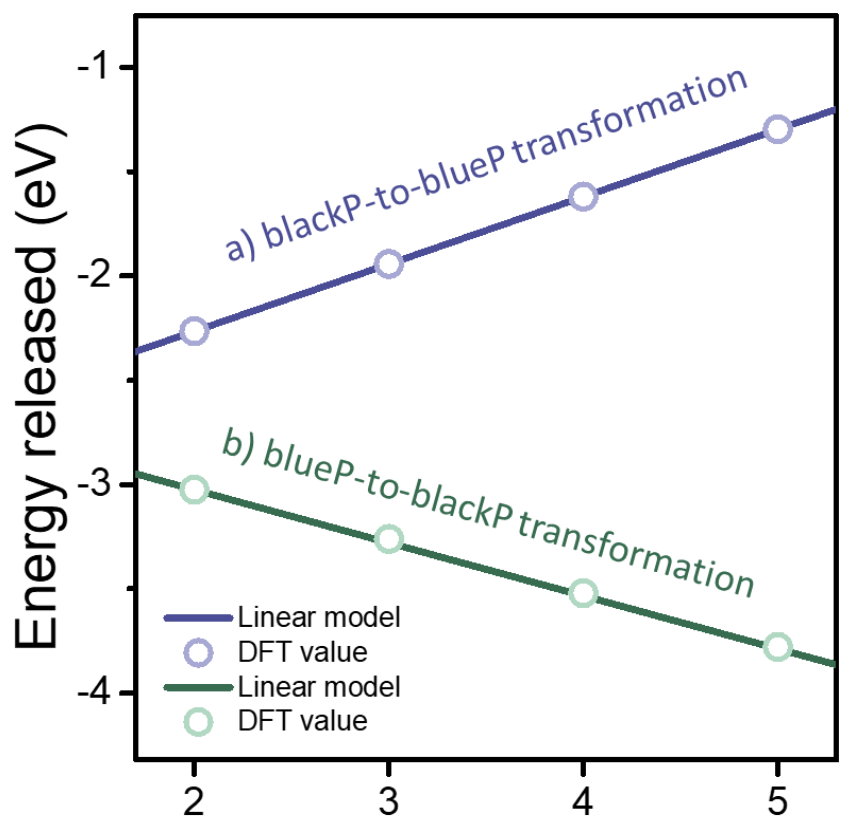

Fig. S8 Comparison of the energy gain (per 8 P atoms) between the linear model and DFT calculations for transitions from (a) blackP to blueP and (b) blueP to blackP.

Assuming allotropes, A and B, with the same stoichiometry are both locally stable. The energy of B is higher than that of A ($E_B > E_A$), so spontaneous transition from A to B is prohibited. However, by applying some external constraints, such as high pressure, the relative energies between phase A and B can be reversed ($E_A > E_B$) and then a transition from A to B becomes favorable. Here we propose that surface chemical adsorption can induce the structural phase transition from A to B. The underlying principle can be explained in a similar way. For phases A and B, where $E_B > E_A$, surface adsorption could revert the energy landscape such that $E_{A'} > E_{B'}$, where $A'$ and $B'$ denote samples with adsorbates. In this way, a surface modification could trigger the structural phase transition or reconstruction of the bulk.

An ideal surface of the layered material A has no dangling bond. Chemical adsorption of foreign atoms (such as Br) on the surface will introduce extra bonds. The new Br-A bond will withdraw (or deposit) electrons from (or to) the surface, leaving un-paired electronic states in the surface region. If any atomic reconstruction, such as layer shifting or interlayer bonding, can stabilize these un-paired electron states, the system will transform to a more stable state accompanied by a reduction of system energy. Such an energy reduction, referred to as $\Delta E_{surf(A')}$, is the key that triggers the phase transition in the deeper layer and eventually the whole bulk region. Note that $\Delta E_{surf(A')}$ does not correspond to the adsorption and dissociation energy of foreign molecules ($E_{ads}$).

To proceed a transition from A to B by surface adsorption, an energy gain ($\Delta E_{A'\to B'} < 0$) has to be achieved. $\Delta E_{A'\to B'}$ consists of the energy differences in the bulk region ($\Delta E_{bulk(A\to B)}$) and in the surface ($\Delta E_{surf(A')}$), which can be written as

$$\Delta E_{A'\to B'} = \Delta E_{bulk(A\to B)} + \Delta E_{surf(A')} \qquad (1)$$

with

$$\Delta E_{bulk(A\to B)} = E_B - E_A = n \cdot (\epsilon_B - \epsilon_A) \qquad (2)$$

where $E_i$ ($i = A, B$) is the total energy of the cell for each phase, $n$ is the number of atoms included in a cell, and $\epsilon_i$ ($i = A, B$) is the energy per atom. Defining $\Delta \epsilon = (\epsilon_B - \epsilon_A)$, one gets

$$\Delta E_{A'\to B'} = n \cdot \Delta \epsilon + \Delta E_{surf(A')} \qquad (3)$$

where both $\Delta \epsilon$ and $\Delta E_{surf(A')}$ are constant. Hence the energy difference is linearly dependent on $n$.

When $\Delta \epsilon > 0$ (which is the case of blackP-to-blueP transformation), one sees from equation (3) that there is a critical value of $n$ of A, less than which a negative value of $\Delta E_{A'\to B'}$ is presented, and the thermodynamics of the transformation is favorable. In our DFT calculations, the linear relationship is reflected excellently as shown in Fig. S8 where the system energy gain is plotted as a function of blackP thickness. Since the energy per P atom in the blueP phase is higher than that in the blackP phase, the slope (Fig. S8a) is positive, and the value of $\Delta \epsilon$, defined in equation (3), is derived to be ~322 meV per 8 atoms. The latter implies that the structural phase transition will be forbidden after a critical thickness, which is estimated to be around 5 layers of blackP (~20 Å) according to the linear model. It is worth mentioning that, if $\Delta \epsilon \leq 0$, there is no 'dead' thickness and phase transition will be more favored for increased thickness. The corresponding results are plotted in Fig. S8b, which again shows excellent consistency with the linear model.

**Layer shift at different pair adsorption coverage**

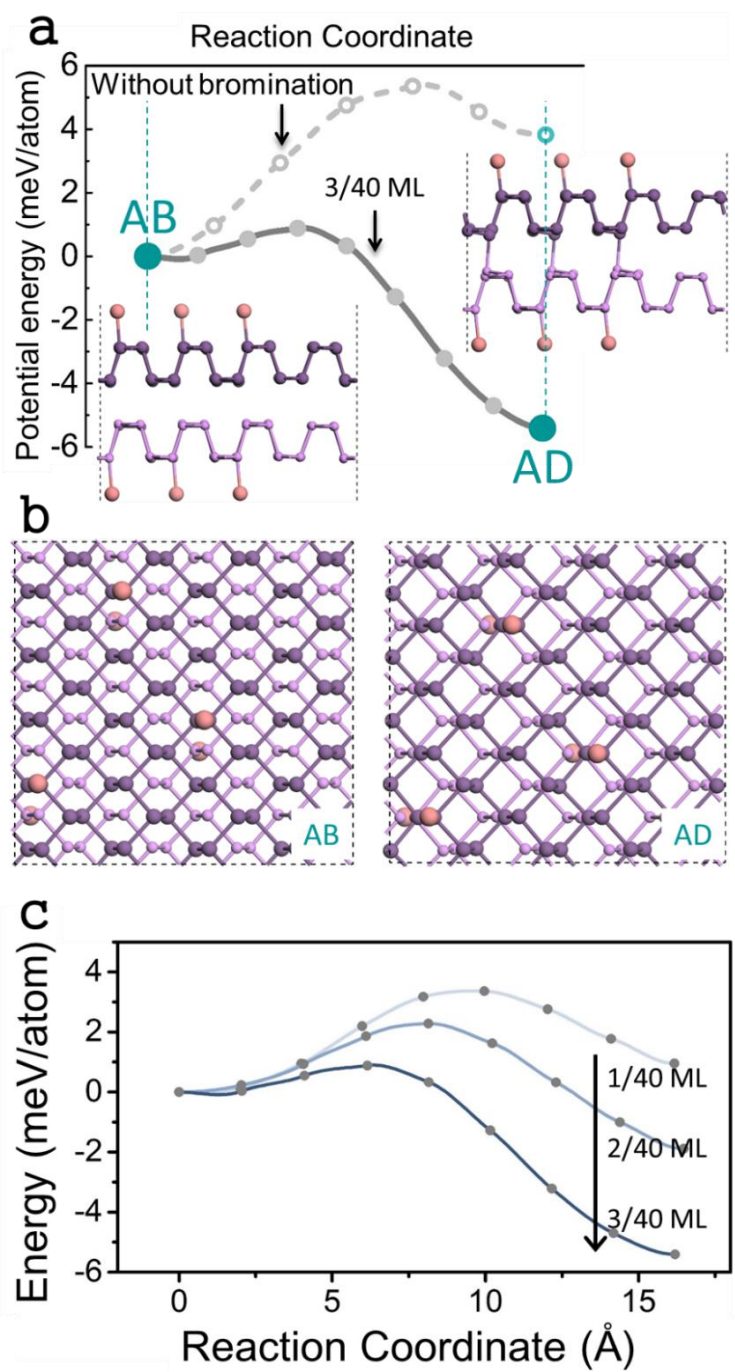

Fig. S10. Comparison of layer shift at different Br coverage. (a) Potential energy for layer shift with 3/40 ML Br adatoms adsorbed on each surface and that without Br adsorption. The inset images show side-views of the AB and AD configurations at 3/40 ML Br coverage. (b) Top view of the AB and AD configurations at 3/40 ML Br coverage. (c) Energy potentials for layer shifts with dual-surfaces Br adsorption of 1/40, 2/40 and 3/40 MLs, respectively.

At the dual-surface Br adsorption of 3/40 ML, the activation energy is found to be 0.88 meV/atom only for two adjacent blackP layers to be shifted from AB to AD stacking (Fig. S10a and b). This shift releases 5.41 meV/atom energy (Fig. S10a). In the absence of Br adatoms, the activation energy is increased by 6.1 times to about 5.35 meV/atom (Fig. S10a). One can see that low levels of bromination significantly influence the sequence of stacking mode of bilayer blackP. According to our calculation, AD stacking prevails once dual-surfaces Br coverage is larger than 2/40 (Fig. S10c).

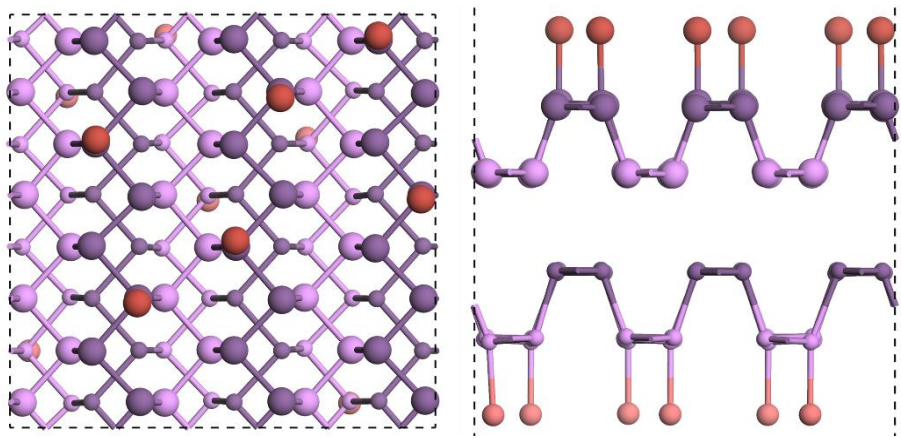
Fig. S10 Top (left) and side (right) views of the RB2 model.

**Kinetics and thermal stability of 2D Br-PNRs**

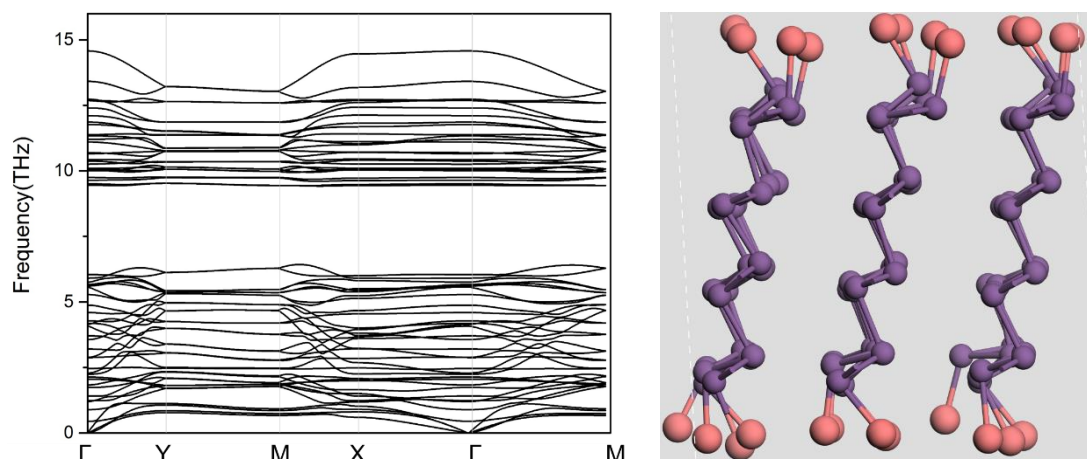

Fig. S11. The phonon modes along the high-symmetry directions (left) and the MD snapshot under 400K annealing for 10 ps of the 2D BrPNRs (right).

The kinetic stability of the layered Br-bluePNRs is confirmed by the absence of imaginary phonon mode in the phonon spectra (see left panel of Fig. S11). Fig. S11 (right panel) also shows the structural snapshot of such Br-bluePNRs after annealing for 10 ps at 400 K, where its atomic configuration remains intact. These results suggest the Br-bluePNRs is both dynamically and thermodynamically stable at room-temperature.

**Self-passivation of 2D BrPNRs**

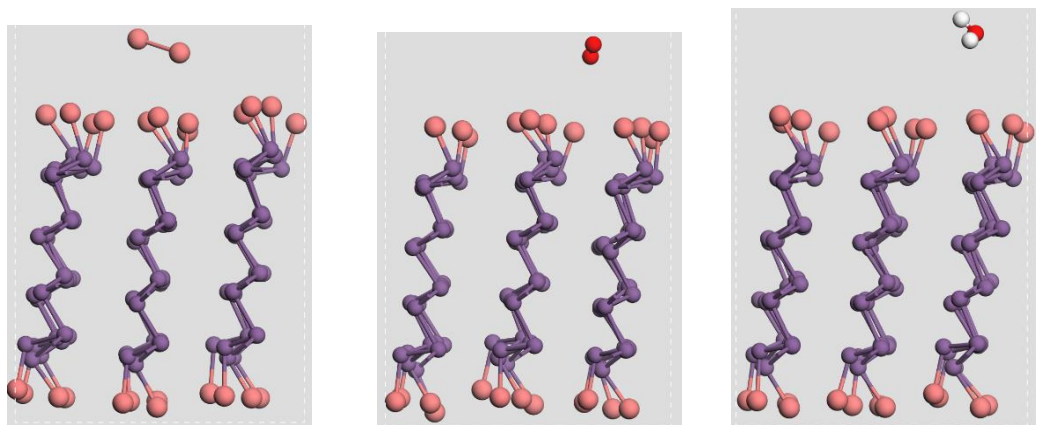

Fig. S12. Snapshot of the structure of 2D BrPNRs in the presence of $Br_2$, $O_2$ and $H_2O$ after annealing at 400 K for 10 ps.

The number of $O_2$ molecule in atmosphere is around $2.7 \times 10^{-5}$ / $Å^3$. One $O_2$ molecule is introduced in a vacuum space of ~2145 $Å^3$, corresponding to twice the concentration in atmosphere. MD simulation is employed to examine the behavior of the $O_2$ molecule at 400K. The same concentration of $H_2O$ is also tested. We find $Br_2$, $O_2$ or $H_2O$ do not interact with the 2D BrPNRs.

Halogenation is reported to endow super hydrophobicity to graphene.[19] The Br-termination could play a similar role as that of oxygen on blackP, where oxygen atoms occupy exposed lone pairs on P and thus prevent further reaction between phosphorus and oxygen.[20] Therefore, compared to the fast degrading blackP caused by the synergy with $O_2$ and $H_2O$,[21-23] BrPNRs probably show long-term stability in ambient environment.

## Lattice parameters and atomic structures of 2-5 layers BrPNRs

Table S1. Lattice parameters of blackP with 2~5 layers and corresponding BrPNRs.

| thickness | blackP a(Å) | blackP b(Å) | Br-PNRs a(Å) | Br-PNRs b(Å) |
|---|---|---|---|---|
| 2 | 4.51 | 3.30 | 4.54 | 6.59 |
| 3 | 4.49 | 3.30 | 4.60 | 6.56 |
| 4 | 4.47 | 3.30 | 4.60 | 6.56 |
| 5 | 4.47 | 3.30 | 4.60 | 6.56 |

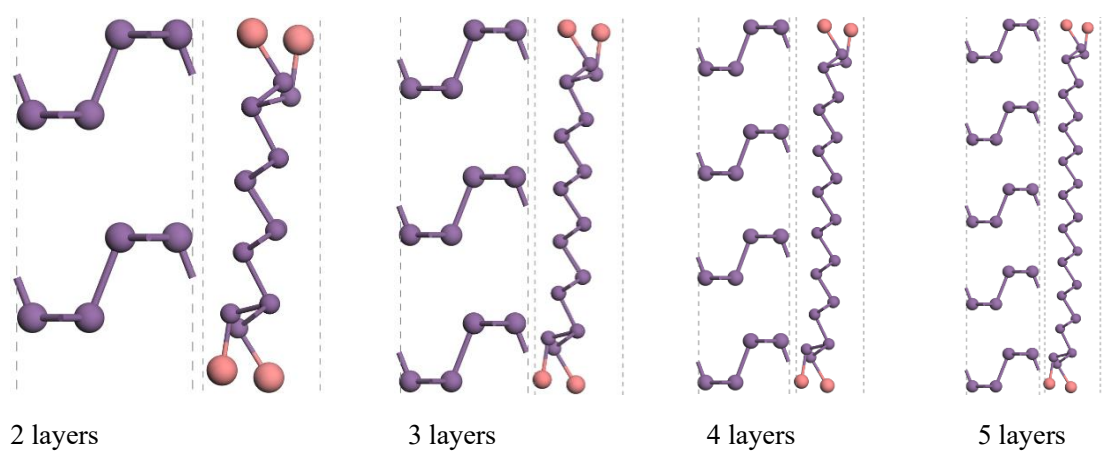

2 layers            3 layers            4 layers            5 layers

The zigzag blue phosphorus nanoribbons with Br terminals (BrPNRs) are stacked into multilayers after the blackP-to-blueP transformation. We assign it as a 2D structure whose rectangular unit cell is spanned in the zigzag and stacking directions. The optimized lattice parameters of blackP with 2~5 layers and the corresponding BrPNRs are shown in Table S1.

Table S2. Carrier mobility predicted by the 1D model.

| Carrier type | $N_L$ | effective masses $m^*/m_0$ | deformation potential (eV) $E_1$ | 1D elastic modulus ($10^{-7}$ Jm$^{-1}$) $C_{1D}$ | mobilities ($10^3$ cm$^2$V$^{-1}$s$^{-1}$) $\mu_{1D}$ |
|---|---|---|---|---|---|
| e | 1 | 0.19 | 5.11 | 0.86 | 0.39 |
| e | 2 | 0.18 | 5.57 | 1.69 | 0.71 |
| e | 3 | 0.18 | 5.74 | 2.44 | 0.99 |
| h | 1 | 3.74 | 1.11 | 0.86 | <0.1 |
| h | 2 | 3.77 | 1.75 | 1.69 | <0.1 |
| h | 3 | 3.11 | 2.23 | 2.44 | <0.1 |